\documentclass[conference]{IEEEtran}
\IEEEoverridecommandlockouts
\usepackage{cite}
\usepackage{amsmath,amssymb,amsfonts}
\usepackage{algorithmic}
\usepackage{graphicx}
\usepackage{textcomp}
\usepackage{xcolor}
\usepackage{comment}
\usepackage{xspace}
\def\BibTeX{{\rm B\kern-.05em{\sc i\kern-.025em b}\kern-.08em
    T\kern-.1667em\lower.7ex\hbox{E}\kern-.125emX}}

\def\Sys{VAMP\xspace}

\def\SysServ{\textit{\Sys Service}\xspace}
\def\ManDb{\textit{Manifest Database}\xspace}
\def\CcfLed{\textit{Distributed Ledger}\xspace}

\begin{document}

\title{Preventing Machine Learning Poisoning Attacks Using  Authentication and Provenance}

\author{\IEEEauthorblockN{Jack W. Stokes}
\IEEEauthorblockA{\textit{Microsoft Research} \\
Redmond, WA USA \\
jstokes@microsoft.com}
\and
\IEEEauthorblockN{Paul England}
\IEEEauthorblockA{\textit{Microsoft Research} \\
Redmond, WA USA \\
pengland@microsoft.com}
\and
\IEEEauthorblockN{Kevin Kane}
\IEEEauthorblockA{\textit{Microsoft Research} \\
Redmond, WA USA \\
kkane@microsoft.com}
}

\maketitle

\begin{abstract}
Recent research has successfully demonstrated new types of data poisoning attacks.
To address this problem, some researchers have proposed both offline and online data
poisoning \textit{detection} defenses which employ machine learning algorithms
to identify such attacks.
In this work, we take a different approach to preventing data poisoning attacks
which relies on cryptographically-based authentication and provenance to ensure
the integrity of the data used to train a machine learning model. The same approach is also used to prevent software poisoning and model poisoning attacks. A software poisoning attack maliciously alters one or more software components used to train a model. Once
the model has been trained it can also be protected against model poisoning
attacks which seek to alter a model's predictions by modifying its
underlying parameters or structure. Finally, an evaluation set or test set
can also be protected to provide evidence if they have been modified by a second data poisoning attack.
To achieve these goals, we propose \Sys which extends the previously proposed AMP system, that was designed  to protect media objects such as images, video files
or audio clips, to the machine learning setting. We first provide requirements for authentication and provenance for a secure machine learning system. Next, we demonstrate how \Sys's manifest meets these requirements to protect a machine learning system's datasets, software components, and models.
\end{abstract} 

\begin{IEEEkeywords}
data poisoning, provenance, cryptographic hash
\end{IEEEkeywords}

\section{Introduction}
\label{sec:intro}

As machine learning models become increasingly ubiquitous within industrial and governmental settings, more effort is needed to maintain, manage and protect the data and software components used to train these models as well as the trained models themselves.
Researchers have recently proposed data poisoning attacks~\cite{Alfeld_Zhu_Barford_2016,li2016data,chen2017targeted,koh2018stronger,wang2018data} where data is specifically altered to achieve some malicious purpose.
In a related attack, which we call a software poisoning attack, the software or the training framework is altered to intentionally introduce a bug or vulnerability. Finally in a model poisoning attack, the model's parameters and its structure may be altered to again produce a malicious output.

Attackers seek to exploit data, software, and model poisoning attacks against machine learning systems during four phases of development and production.
First, the training and validation datasets that are used for training a model may be altered in a data poisoning attack either by insider threats or from man-the-middle attacks which modify the data during transmission over a network such as the Internet. For example, a face recognition model which is trained using poisoned data may provide attackers with a backdoor into computer systems allowing an unauthorized user to pose as a valid user~\cite{8953400}. Second in a software poisoning attack, the machine learning software, packages, or containers (e.g., Docker)  used to train the model may be maliciously altered to introduce a vulnerability. Examples of software poisoning attacks include the introduction of a difficult-to-discover bug, malware or a backdoor into the machine learning system infrastructure. Third, if the model was trained with pristine (i.e., clean) data and software, it may be modified in a model poisoning attack to produce incorrect results during inference. Finally, it may be possible to conduct an additional data poisoning attack against the unknown data used for inference or the test dataset used to evaluate the model's performance. As a result, we seek to prevent all types of attacks against the data, software, or the trained models.

One method for preventing data poisoning attacks is data poisoning detection which falls into two main approaches: offline~\cite{10.1007/978-3-030-00470-5_13} and online~\cite{wang2020practical}. In an offline detection system, algorithms analyze the trained model to determine if it was trained with poisoned data. In online detection, algorithms seek to monitor the data being used to train the model during the training process and identify poisoned data. Since data poisoning detection relies on statistical algorithms to detect these types of attacks, they produce both false positives and false negatives.

Provenance has been previously proposed to protect machine learning systems~\cite{10.1145/3128572.3140450,8473440}. In this work, we take a different approach to the problem of preventing data poisoning attacks. We propose the use of \textit{cryptographic hashing}, in addition to provenance, in order to protect the original datasets that are used for training and validation. In addition, cryptographic hashing also ensures the integrity of the software used to train or evaluate the model. When a model has been trained, we again use cryptographic hashing in order to ensure that it is not been tampered with by attackers. This allows us to prevent the incorrect use of the model during inference. If the cryptographic hashes ensure that the datasets, software, and model have been authenticated using cryptographic hashes, the system's integrity is assured.

Using provenance and cryptographic hashing to combat fake media, including photoshopped images and both cheapfake and deepfake videos, has been recently proposed~\cite{england2020amp,Origin,CAI2}. A number of examples of this approach include Project Origin~\cite{Origin} and the Content Authenticity Initiative (CAI)~\cite{CAI2}. CAI focuses on images captured in a camera, using a secure hardware enclave, through the content creation process using photo editing tools~\cite{CAI2}. Project Origin, which is an alliance between the BBC, CBC, Microsoft and the New York Times, instead protects the integrity of images and videos from the point of initial publication to the display on a webpage or in an app~\cite{england2020amp,Origin}.

Project Origin uses the AMP system~\cite{england2020amp} as the underlying authenticity and provenance technology. AMP is a proof-of-concept Azure web service which includes several components that combine to convey provenance to the end user when consuming a piece of media. A publisher first creates metadata related to the image, video, or audio clip which embeds this metadata in a data structure we call a manifest, cryptographically binds the manifest to a media object via object hashes, and then signs the manifest. The publisher then uploads the manifest to the manifest database in the web service or alternatively embeds the manifest into the media file itself. Next, the manifest or some subset of the manifest can be optionally inserted into an immutable, distributed ledger using the Confidential Consortium Framework (CCF)~\cite{CcfCode,CcfDoc,CcfTech}. This distributed ledger provides publicly available evidence that the manifest has not been modified by attackers.

In this paper, we extend AMP to create a new authenticity and provenance system called \Sys which aims to prevent data, software and model poisoning attacks aimed at all aspects of the machine learning system. We first define the requirements of a cryptographically-protected machine learning provenance system. Next, we show how \Sys fulfills these requirements with only minor modifications to the underlying AMP system.

The main contributions of this paper include:
\begin{itemize}
  \item We propose a cryptographic-based authentication and provenance solution for machine learning systems for the prevention of data, software, and model poisoning attacks against the training, validation, and test datasets, the machine learning software and components, and the trained model.
  \item We list the requirements for different phases of developing and deploying a machine learning system that need to be addressed by a machine learning authentication and provenance system.
  \item We extend the AMP system to the machine learning setting and show how its relational concepts that are important for protecting media are also important for machine learning.
  \item We use a highly performant distributed ledger using the confidential consortium framework to offer public assurance of the integrity of the objects in the machine learning environment.
  \item We implement this system as an Azure web service.
\end{itemize}

\section{AMP}
\label{sec:amp}
\begin{figure}[tb]
\centering
\includegraphics[trim = 1.0in 3.0in 0.2in 3.0in,clip,width=1.0\columnwidth]{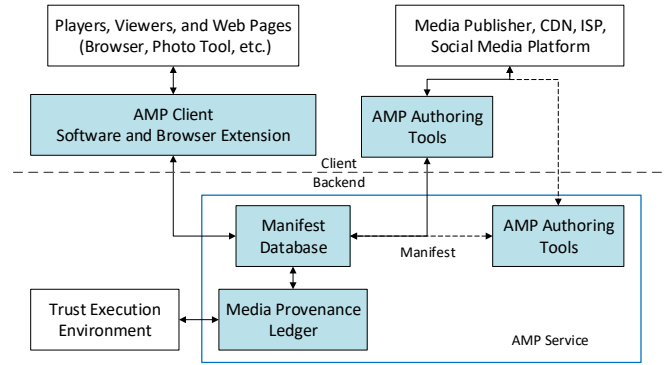}
\caption{High-level overview of the AMP system components (shaded boxes).}
\label{fig:system}
\end{figure}

\Sys can be considered as an extension of the AMP system~\cite{england2020amp} which provides cryptographically-authenticated provenance for media. In this section, we briefly review the \textit{AMP Service} which is
depicted in Figure~\ref{fig:system}.
Its key components include a manifest which provides the metadata for the media objects, the \ManDb which stores previously uploaded manifests, and a \CcfLed (i.e., blockchain) which provides publicly available evidence that the media has not been modified. In addition, the manifest includes fields which allow the media consumer to track its provenance backwards through the media capture, editing, publishing, and distribution graph to its original source.
%

AMP manifests can be stored using two separate methods, embedded or detached. AMP proposes a new way of embedding (i.e., inserting) manifests
into MP4 files. In the detached manifest scenario, the manifest is instead stored separately in the \ManDb. A client application such as a browser extension or a webpage then can authenticate media using the file itself if the manifest is embedded or using the \textit{AMP Service} if the manifest is detached.

AMP serializes the manifest in two ways using both CBOR and JSON. CBOR is more efficient since it
stores the data in a binary format whereas a manifest stored using JSON in a plaintext format is human readable.

AMP uses a X.509 PKI (public key infrastructure) trust model. The serialized manifests are signed with the publisher's private key allowing client applications to verify the publisher's identity.

In following sections, we discuss how the media-focused AMP system can also be extended to help protect machine learning systems.

\section{Threat Model}
\label{sec:threat}
\Sys's threat model is similar to that for AMP. 
All of \Sys's computing infrastructure, including every computer's host operating system (OS), the \SysServ, the \ManDb, and the software tools, are assumed to be secure. The \CcfLed and its trusted execution environment are also assumed to be secure. In addition, the X.509-based trust environment, including the certificate authority (CA) infrastructure and cryptographic hashing algorithms, are assumed to be secure.
These assumptions are reasonable because the VAMP system is designed, built and managed with security as the primary consideration.

Conversely, machine learning datasets and even some computing systems, do not have security built in from the ground up. Training data is widely replicated but weakly authenticated (rarely more than an MD5 hash on a website) and so is especially vulnerable to adversarial changes. Similarly, open source code is frequently forked and locating the legitimate version may be problematic. Software packages, and their dependencies, are frequently updated by unknown individuals. The resultant trained models are similarly redistributed with no or poor authentication.
\section{Authentication and Provenance of Machine Learning Systems}
\label{sec:mlprov}
Both authentication and provenance play critical functions in a \textit{secure} machine learning system. Authentication of machine learning system's components such as datasets, software and models ensures the veracity of the final results. Similarly, provenance  allows the model trainer, and to some extent the user, to trace and verify all of the components that were used to train the system backwards through the provenance graph to the original sources.

\subsection{Authentication}
In a secure machine learning system, authentication is the process of the consumer verifying the veracity all of the machine learning objects by first verifying the creator's signature for each object and then verifying that each object's one or more cryptographic hashes were generated from its contents.
Before using a machine learning object in a secure system, the object consumer first verifies the signature of the object. 
Building a cryptographically-protected provenance system for the prevention of data, software, and model poisoning attacks
involves three main tasks including protecting the training and validation datasets, protecting the software used to train and evaluate the model, and
later protecting the trained model so that it can be used for tamper-evident inference. In addition, the evaluation dataset
can also be protected and verified before inference as an optional fourth task.

\begin{figure*}[tb]
\centering
\includegraphics[trim = 0.0in 0.0in 0.0in 0.0in,clip,width=0.7\linewidth]{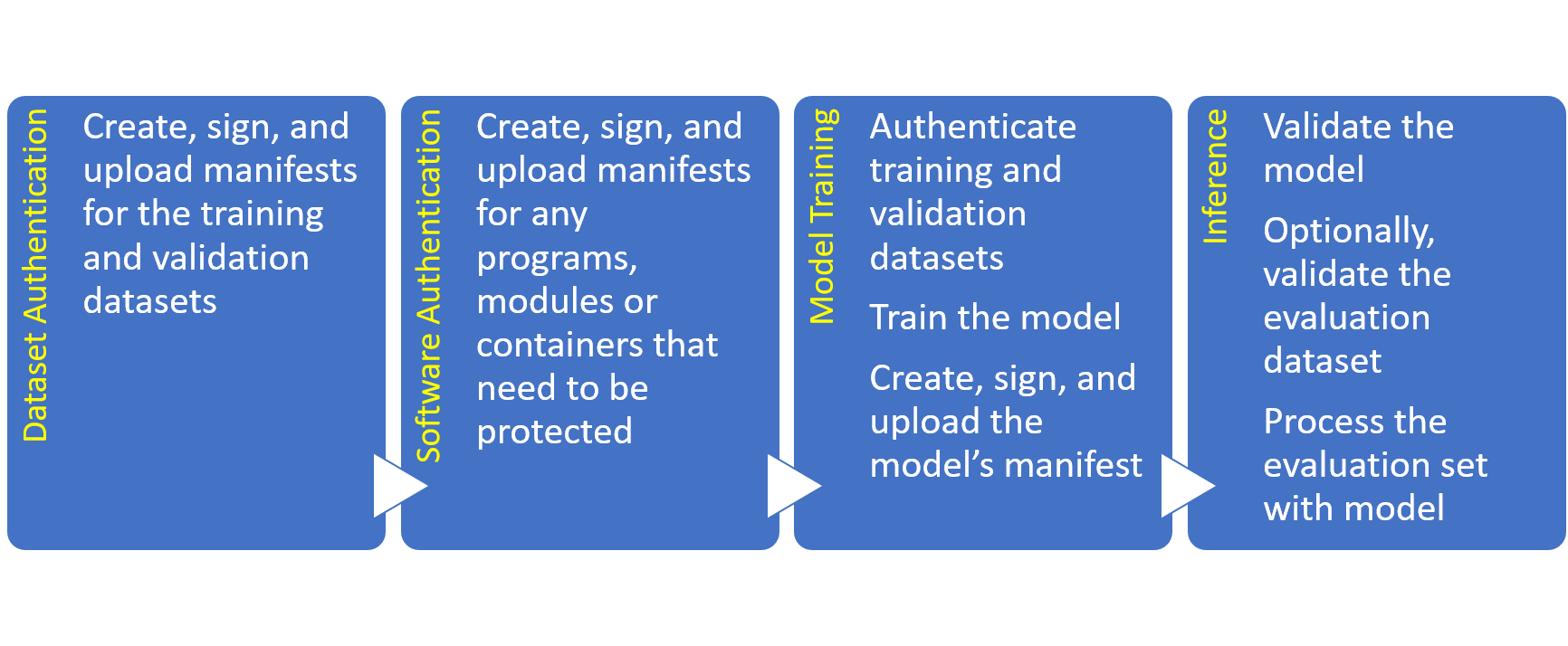}
\caption{Provenance steps for training and evaluating datasets and a machine learning model.}
\label{fig:flow}
\end{figure*}

\Sys's pipeline in illustrated in Figure~\ref{fig:flow}. First, the dataset creator generates the
training and validation datasets. Next, they create the manifests (i.e., metadata and data bindings) and either upload (i.e., publish)
them to the \Sys service or embed them directly in the individual datasets.

All of the software (e.g., source code, packages, containers) that is used for training the model as well as any software that is used to evaluate the model is then uploaded to the service.

Next, the model
creator validates that the content bindings in the manifests match the data and labels in the dataset and the model training software components.
After the validation succeeds, the model creator trains the machine learning model. Afterwards,
the model creator generates a manifest for the model and either uploads the manifest to the \Sys
service, or embeds the manifest into the model itself. 

Once the model has been trained and its manifest uploaded
to the service, the user can perform inference on the unknown evaluation set.
Similar to the previous steps in this
process, provenance is also an important aspect for inference.
The user first validates that the model bindings match the contents of the model file. It is possible
at this point that the manifest for the evaluation dataset has been previously generated and either
uploaded to the \Sys service or embedded in the dataset. If so, the user also validates that
the bindings in the manifest match the data, and labels if they exist, in the dataset. After validating the model,
and possibly the evaluation dataset, the user produce the final scores for the evaluation dataset using the
model.

\subsection{Provenance}
In a machine learning system, provenance involves being able to understand how the all of the different machine learning objects are processed to yield the final prediction score. Provenance also plays a key role in securing a machine learning system. Provenance can  typically be represented as a directed acyclic graph (DAG), and therefore provenance can be traced in either direction from the start to the  finish or backwards from the end to the beginning. Many complex machine learning systems consist of simpler machine learning components. Each of these components may be trained with different datasets which may or may not be standard (e.g., ImageNet). In order to verify that the final complex model has been trained using authenticated datasets, software, and underlying (i.e., upstream) models, it requires a provenance graph which indicates all of these system components. Provenance is also important to creating machine learning systems that generate reproducible results~\cite{paganini2020dagger,samuel2020machine}. 

\subsection{System Overview}
\begin{figure}[tb]
\centering
\includegraphics[trim = 1.0in 3.0in 0.2in 3.0in,clip,width=1.0\columnwidth]{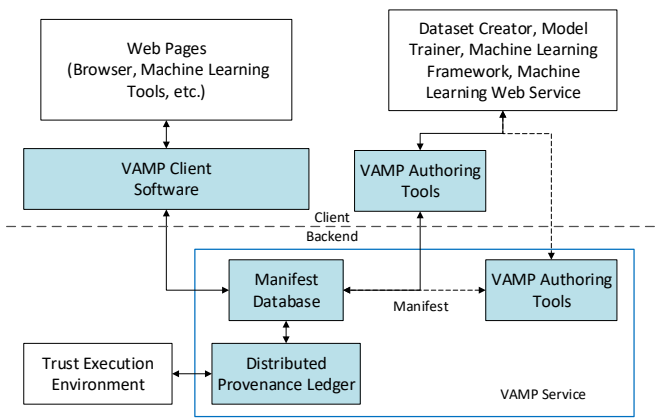}
\caption{High-level overview of the VAMP system components (shaded boxes).}
\label{fig:vamp_system}
\end{figure}

Figure~\ref{fig:vamp_system} depicts the \Sys system
which has been extended from the AMP system shown in 
Figure~\ref{fig:system} for the machine learning use case.
\Sys provides authoring tools for creating the manifests locally and a web app for creating the manifests in the service. 

In the next sections, we first describe the requirements imposed by many machine learning systems on datasets, software components,
and the training process. We then describe how manifests and \Sys can be meet these machine learning requirements. 
\section{Datasets}
\label{sec:dataset}
The first step to preventing data poisoning attacks in machine learning systems
is creating datasets with manifests which cryptographically bind the metadata to the data and labels, if they exist.

Since data poisoning attacks primarily target the model training process.
It is most important to protect the training and validation datasets
so the model training code (e.g., Pytorch, Tensorflow, ML.Net) can authenticate the
data used during training.

Unlike formatted media objects which are addressed by AMP, machine learning
datasets are often text files, although the raw data underlying vision 
datasets are standard image formats such as JPEG. Text file-based machine learning
datasets typically have a custom format which either has a prescribed format definition
or includes a header which defines this format. It may not be possible to modify standard datasets since
the metadata cannot be inserted into the dataset without breaking existing training code.
In this case, the dataset creator's manifest must be stored externally such as in a separate file or in a
web service. For new datasets, the metadata format could be specified 
and inserted into the text file itself.  

Another important requirement of a secure machine learning system is determining what is the important metadata
needed for the datasets and what are
important aspects (i.e., fields) of the metadata to cryptographically protect.

\section{Machine Learning Software}
\label{sec:code}
In addition to protecting the datasets, it is also important to protect all of the software used to train the model to prevent software poisoning attacks.
For machine learning systems, this software can either be text-based software such as Python, Java, or C\# code or it may be binary packages needed to create features (e.g., OpenCV) or train the model itself.
Machine learning models are often trained within containers (e.g., Docker) and these containers need to be protected as well.
Thus, it is important for a machine learning system to protect all aspects of the machine learning software training and inference environments. 
\section{Model Training}
\label{sec:training}
After the manifests for the training and validation datasets and all software components have been successfully uploaded to the authentication service or embedded into the files,
the datasets, software and manifests can be used to train the machine learning model with cryptographic guarantees that the components have not been altered.
The software must be validated before training the model.
The datasets can either be verified at the start of training or continuously verified during the training process if the dataset is provided 
using a streaming service. After training has completed, it is important to protect the final model used to evaluate unknown datasets during inference.

\subsection{Verifying the Datasets Before Training}
In the majority of cases, the training and validation datasets can be verified once before training the model. This verification
may be done by computing a single hash for each dataset or the verification may be done in chunks for more efficient, and perhaps parallel, verification.

\subsection{Multiple Minibatch Sizes}
When training standard machine learning models such as logistic regression, the parameters are updated based on the loss for each \textit{individual}
sample in the training set. In this setting, the order that the samples are processed is chosen at random for each subsequent epoch.
On the other hand, deep learning models are typically trained using minibatches, and one important hyperparameter that is chosen during 
training is the minibatch size.  Instead of randomly choosing the order for each sample as done in the standard machine learning setting, 
the order of contiguous minibatches is chosen randomly when training deep learning models.

\subsection{Adaptive Minibatch Sizes}
A recently proposed deep learning training approach seeks to learn the optimal minibatch size during training~\cite{alfarra2020adaptive}. 
For adaptive minibatch size training, the objective function includes a term which
allows the minibatch size to be changed during training using stochastic gradient descent (SGD). Thus, if adaptive minibatch size training is required, the secure machine learning system must be able to support dynamically changing the minibatch size from epoch to epoch.

\section{\Sys Manifests}
\label{sec:manifest}
Next, we demonstrate how \Sys fulfills the requirements of a cryptographically-protected, machine learning provenance system.
The manifest is the key data structure in \Sys. Its main two functions are to 1) define and cryptographically protect the critical metadata that is important for the datasets, software, and models, and 2) cryptographically bind this metadata to these machine learning objects. In addition, manifests can also define relationships between these
objects which allow the model trainer, and maybe even the user in some cases, to trace back the machine learning object provenance from the final prediction score or the inference object back to all of its original components.

\subsection{Key Manifest Metadata Fields}
All of the metadata structures related to media authentication and provenance are described in the AMP system design and are provided in Appendix C in~\cite{england2020amp}.
We believe that the AMP metadata structures can also be used for machine learning metadata with minor modifications.
Thus, we have not added or removed any of metadata fields for \Sys.

The key manifest fields are listed in Table~\ref{tab:static_man}. These fields match those originally specified in AMP with one exception. AMP specified a MediaID, but this field has been changed to ObjectID in \Sys allowing it to generalize to the machine learning scenario. The ObjectID is the identifier of the machine learning object that is being protected.
In addition, a number of the AMP field descriptions reference media related concepts. In \Sys, these field descriptions are updated to be more general.

\begin{table*}
  \begin{center}
      \begin{tabular}{|l|l|l|}
        \hline
        Field & Manifest Type & Description \\
        \hline
        \hline
        ObjectID & Static/Streaming & Publisher-assigned identifier for the object.  \\
        \hline
        MasterCopyLocator & Static/Streaming & URI of a stable, publisher provided location service or a generic URL redirector service.\\
        \hline
        EncodingInformation & Static/Streaming & String describing the object type (e.g., ``JPEG'', ``MP4'', ``Gzip'', ``Huffman encoding'', ``Run-length encoding''). \\
        \hline
        OriginManifestID[] & Static/Streaming &  One or more ManifestIDs that describe the source object used to create a derived work.  \\
        \hline
        Copyright & Static/Streaming & Copyright string associated with the object.\\
        \hline
        ObjectHash[] & Static & Cryptographic hash of the associated simple object (or collection of related objects \\
        & & objects). \\
        ChunkDigest & Streaming & An ordered array of chunk-hashes starting from the beginning of the work. \\
        \hline
      \end{tabular}
  \end{center}
  \caption{Key manifest fields. 
           }
  \label{tab:static_man}
\end{table*}

\subsection{Manifest Data Binding}
Manifests are crytographically bound to the data, and the labels if they exist, located in the datasets by computing
one or more cryptographic hashes of the data and labels. Similarly, the cryptographic hashes are also computed for any software components and the final trained model. \Sys uses SHA2 cryptographic hashes such as SHA2-256 or SHA2-512 for this purpose.

\Sys supports three different types of content binding: static, fixed-length chunk, and box-based. For a static content binding, a cryptographic hash is computed over the entire contents of the object (e.g., machine learning small dataset, source code, model) that need to be protected. For large datasets using a fixed-length chunk content binding, the file is divided into consecutive fixed-length chunks, and the cryptographic hash is computed for each chunk.

In AMP, the box-based binding is used to insert the metadata and content hashes into the media file according to its format specification, and this type of binding can also be extended to support the protection of machine learning objects.
\Sys's layout of a machine learning object is provided in Table~\ref{tab:vamp_text_file} for files where the header can be modified. This format supports both embedded and detached manifests. The beginning of the file contains the ManifestType which indicates if the manifest is embedded or detached. The next field, ManifestSerialization, specifies if the manifest is stored in canonicalized JSON or canonicalized CBOR. If the ManifestType indicates that the manifest is detached, the next field ManifestLocator provides the URI of the manifest. Otherwise, the signed manifest is embedded and located in the next section of the file. For text files, the manifest is Base64 encoded. Finally, the media object itself is stored in the remainder of the file.

The box-based binding can also be used to support datasets with minibatches of different lengths for training deep learning models. Since the best minibatch size is typically learned during hyperparameter tuning, it may not be possible to set this parameter beforehand. Therefore, the challenge for supporting different minibatch sizes is that the minibatch hashes must computed for each possible setting of the minibatch size hyperparameter.

An example of a text-based dataset with an embedded manifest is depicted in Table~\ref{tab:vamp_dataset_file}.
where the data is divided into minibatches $1$ though $N$. These minibatches contain a fixed number of examples (e.g., 64). However since each row in the dataset consists of a potentially different number of characters, each minibatch can have a different length and be located at a non-uniform offset from the beginning of the file. The box-based binding is also important in the case where the user may want to repeatedly validate each minibatch for an epoch during training. Therefore, AMP's box-based binding format has been extended in \Sys to  handle chunks of varying offsets and lengths to accommodate minibatches in a deep learning dataset.

Machine learning datasets can also be stored in a binary format instead of a text file. In this case, the minibatches are contained within a fixed number of bytes. However, these fixed-length minibatches may still be stored with a random offset due to the inclusion of the manifest and the column specification header, and the box-based content bindings are still required.

\begin{table}
  \begin{center}
      \begin{tabular}{|l|l|l|}
        \hline
        Field & Storage & Description \\
        \hline
        \hline
        ManifestType & Embedded / & Specifies if the manifest  \\
        & Detached & is Embedded or Detached.  \\
        \hline
        ManifestSerialization & Embedded / & Specifies if the manifest  is \\
        & Detached & serialized using \\
         &  & canonicalized JSON  \\
                  &  & or CBOR.  \\
        \hline
        ManifestLocator  & Detached & URI of the manifest.\\
        \hline
        Manifest & Embedded & The signed manifest \\
        & &  of the machine, \\
        &  &  learning object. \\
        \hline
        Data/Code  & Embedded / & The data for a dataset, \\
        & Detached & the code for a program., \\
        & & the binary data for a sofware  \\
        & & component, or the architecture   \\
        & & and parameters for a model. \\
        \hline
      \end{tabular}
  \end{center}
  \caption{Media object file structure supporting both detached and embedded manifests.}
  \label{tab:vamp_text_file}
\end{table}

\begin{table}
  \begin{center}
      \begin{tabular}{|l|l|l|}
        \hline
        Field & Description \\
        \hline
        \hline
        ManifestType  & Specifies that the manifest  \\
        & is Embedded.  \\
        \hline
        ManifestSerialization  & Specifies if the manifest  is \\
         & serialized using \\
         &  canonicalized JSON  \\
                  &  or CBOR.  \\
        \hline
        Manifest & The signed \\
        & manifest of the dataset. \\
        \hline
        Data &  Minibatch 1 \\
         & \\
        \hline
        Data & Minibatch 2 \\
        & \\
        \hline
        Data & ... \\
        \hline
        Data  & Minibatch N \\
         & \\
        \hline
      \end{tabular}
  \end{center}
  \caption{Text-based, dataset file example with an embedded manifest and fixed-length minibatches.}
  \label{tab:vamp_dataset_file}
\end{table}

Once the hashes have been computed and inserted into the manifest, along with the other important metadata, the key parts of the manifest that need to be protected are then
signed to provide evidence if an attacker has modified the dataset, software or model.

\subsection{Transformation Manifests}
Transformation Manifests are used to indicate provenance in a machine learning system. All of the datasets, software components, and models can form a complicated graph. Understanding this complete graph is important for implementing reproducible machine learning systems. Transformation Manifests include one or more backpointers to the manifests of other machine learning objects. These backpointers support a derived object relationship and can be used in a number of ways.
For example, one common practice in machine learning is to start with a pretrained model which has been trained with data and labels for another purpose. Using this pretrained
model as a starting point, it is finetuned with a different set of data and labels to achieve another objective. In this case, the finetuned model is ``transformed'' from
the original, pretrained model, and a backpointer from the finetuned model to the pretrained model indicates this relationship.

In another example, an uncompressed dataset can be ``transformed'' from a compressed dataset.
Decoding an encoded dataset into a text or binary file can be represented using a Transformation Manifest. In addition, an encoded dataset may be transcoded from one lossless compression algorithm to another
including compression algorithms such as Gzip, Huffman encoding, or Run-length encoding, and this transcoding operation can be represented using another Transformation Manifest.

A Transformation Manifest also supports the notion of derivation from multiple objects. In machine learning, we can create a Transformation Manifest which contains backpointers to the manifests of the training and validation datasets used to train a model, for example. In another case, the Transformation Manifest may provide pointers back to multiple submodels used during preprocessing steps to create features for the final model.

\subsection{Facsimiles}
Facsimiles are datasets, and possibly even models, that the data publisher or the model creator believe are similar or related in some way.
Manifests can authenticate any number of facsimiles in
machine learning scenarios where facsimiles may be used for:
\begin{itemize}
  \item Splitting one labeled dataset into separate training, validation, and test datasets
  \item Training with different minibatch sizes
  \item Adaptively selecting the minibatch size during training
  \item Constructing  datasets of different sizes, including subsampling and oversampling, from a single dataset.
\end{itemize}

In the first example, a single dataset can be split into separate training, validation, and test datasets.
In this case, the single large dataset can be considered to be a facsimile of a collection of the individual training, validation, and test datasets.
Another example of a pair of datasets which can be considered as facsimiles includes
two datasets where the first is a single multiplexed dataset containing both the data and labels,
and the second includes two datasets where the data and the labels are stored separately.

Switching the minibatch size during adaptive minibatch training is similar to the concept of adaptive bitrate streaming which is commonly
used to adjust the video quality of internet-streaming video due to changes in the network.

\subsection{Manifest Storage Locations}
In the original design, AMP allows for manifests to differ based on their storage location. The different storage types of AMP manifests include  detached and embedded. Likewise, \Sys includes both detached and embedded manifests. Furthermore, since machine learning datasets are typically stored in text-based files, as opposed to well known binary media format files (e.g., JPEG, MP4),  \Sys further allows for \textit{detached} manifests to be stored locally or in the cloud. Finally, \Sys also allows both detached or embedded manifests to be stored both locally \textit{and} in the cloud simultaneously.
\begin{itemize}
  \item \textit{Detached Manifests:} Detached Manifests are manifests which are stored separately
from the machine learning objects themselves. Datasets are not implemented using standard structured text or binary formats,
and similarly, current machine learning models trained with PyTorch or Tensorflow have a prescribed
model format. In both of these cases, the manifest can be created and uploaded to the \SysServ after
the dataset has been created or the model has been trained.

Detached Manifests can either be \textit{Detached Local Manifests}
or \textit{Detached Cloud Manifests}. Detached Local Manifests are stored in the same directory structure (e.g., in the same
directory) whereas Detached Cloud Manifests are stored in the cloud and uploaded or read using a web service.
  \item \textit{Embedded Manifests:} Embedded Manifests are inserted into the media object files which allows them to be easily authenticated. Unlike media
which relies on standard encoding formats (e.g., JPEG, MP4), machine learning datasets are not typically stored using a standard format.
On the other hand, machine learning models are usually stored using a format defined by the machine learning
framework (Pytorch, Tensorflow, Scikit-Learn).
We anticipate that as cryptographic-based provenance solutions become more ubiquitous, it is possible
that a current dataset or model structure
can be extended, or new structures can be created, to allow the inclusion of a manifest.
  \item \textit{Manifests Stored in Multiple Locations:} This case includes storing the manifest
in two locations, locally and in the cloud. For example, a media object's manifest can be stored in a Local Detached Manifest file in addition to being stored as a Detached Cloud Manifest in the \ManDb. Since media is typically stored as self-contained \textit{binary} files, AMP only supported \textit{Embedded Manifests} or
\textit{Detached Cloud Manifests}, where embedding manifests in the file is the preferred method for media. Thus, the original AMP design did not consider storing the manifest in both locations.
\end{itemize}

\subsection{Detached Local Manifests}
As noted above, a machine learning system may consist of a number of different machine learning related files such as the source code for training or inference, separate
training, validation, and test datasets and the trained model itself. Furthermore, each dataset may be separated into two different
files for the data and the labels. As such, a directory structure may contain many different \textit{Detached Local Manifest}
files. If the orignal media object file can be modified, a reference to the manifest file can be included using the ManifestLocator in Table~\ref{tab:vamp_text_file}. However, it may not be possible to add this additional metadata to the original media object file. In this case to allow the user to quickly associate the manifest file and the machine learning object file when storing the manifest
as a \textit{Detached Local Manifest}, \Sys requires the following naming convention. If a machine learning object is stored
as \textit{name.xxx}, then the manifest is stored in the same directory as \textit{name.xxx.man}. For example, the training set might be stored
as \textit{data/training.csv}, and the manifest would then be stored as \textit{data/training.csv.man}.

\subsection{Serialization}
As noted in Section~\ref{sec:amp}, AMP serializes manifests in two ways using both canonicalized JSON and canonicalized CBOR. Since media formats are binary, using
the CBOR binary serialization format for manifests is a natural fit. However for text-based datasets typically encountered in machine learning
systems, CBOR serialization for Embedded Manifests would result in binary data being inserted into a file which otherwise consists of text.
Thus, \Sys allows for the option of serializing the manifests using either JSON for text-based datasets or CBOR for datasets implemented in a binary format. Binary arrays are Base64 encoded before serialization.

Manifests are signed by the media object creator to insure that the machine learning object has not been modified
from its original version. Canonicalized JSON manifests are signed using a JWT signature block while canonicalized CBOR manifests are signed
using COSE signature blocks.

\subsection{Verification}
Before training or inference, the signed manifests of each required machine learning object must be first verified. For training, the manifests  for the training and validation datasets and the training software components are verified using the signer's public key. If the media object is serialized using COSE, the manifest is Base64 decoded to reveal its authenticated contents. The signer's public key may be different for each machine learning object. Similarly, the dataset, software components and model can also be verified during inference using the same process.

\section{Trust Model}
\label{sec:trust}
As mentioned previously, AMP uses X.509 to implement its trust model. Similarly, \Sys also employees X.509 and certificate authorities (CAs) to establish the authenticity of machine learning datasets, software and models. A dataset creator or a model trainer's certificate is used to sign 
the machine learning object's manifest in \Sys. The certificates can be issued by any standard CA.

In general, determining the identity of the entity who signed
a media object is much easier for \Sys users than for AMP users who are trying to confirm the identity of the person or organization who published a media object. The main reason is because machine learning objects are much less prevalent than all of the media objects that are available on the Internet.

We have not identified any changes needed for AMP's trust model to enable the machine learning scenario addressed by \Sys.

\section{Distributed Ledger}
\label{sec:ledger}
A distributed ledger provides additional security guarantees over a simple signature. AMP uses a distributed ledger to further ensure the veracity of media, and this ledger is implemented using the Confidential Consortium Framework (CCF)~\cite{CcfDoc,CcfCode}.
Since \Sys is an extension of AMP, it also employs CCF to help ensure the integrity of the machine learning system by providing a public audit trail of the media objects that were either used to train the system or to evaluate new data.  One particular advantage is that the ledger preserves the order of publication of different objects: digital signatures require a local clock (which may be adversarial set) or need to use a trusted timestamp authority.

In the case where the manifest for the media object is uploaded to the cloud service, the entire manifest or its cryptographic hash would also be written to the CCF ledger. CCF provides a receipt which can be used to ensure the authenticity of the machine learning object without the need to query the ledger itself. Examples of the performance for storing the manifest in the ledger are provided in~\cite{england2020amp} for a number of different Azure data center configurations.
\section{Implementation}
\label{sec:implementation}
\Sys is primarily implemented in C\#. CCF is implemented in C++. To test \Sys, we also use C\# with ML.NET test cases. The Azure web service is implemented also written in C\#. Webpages for creating manifests are implemented in C\# using Razor. 
\section{Related Work}
\label{sec:related}
Provenance solutions have been previously proposed for data and machine learning systems.  
One early work which frames the need for data provenance is~\cite{10.1007/3-540-44503-X_20}.
A blockchain solution was proposed for data provenance in the IoT setting~\cite{10.1007/978-3-319-68136-8_8}.
Provenance systems~\cite{10.1145/3128572.3140450,8473440} have been previously proposed for machine learning systems against data poisoning attacks. However unlike \Sys, neither system used cryptographic hashes to ensure the integrity of the data and software. 

\Sys is also related to creating reproducible machine learning systems, debugging machine learning systems and incorporating explainability in machine learning systems.
One example of a reproducible machine learning framework is dagger~\cite{paganini2020dagger}. Provenance has been used for other reproducible machine learning environments including~\cite{souza2019provenance,xin2021production}. Principles related to provenance, reproduciblity and FAIR data are given in~\cite{samuel2020machine}. Another framework for debugging machine learning system is given in~\cite{Louren_o_2019}. An interactive and explainable framework for machine learning is proposed in~\cite{8807299}.

Provenance systems have also been proposed for authenticating media. 
Since \Sys is an extension of AMP, it is most similar to that system~\cite{england2020amp}. In this work, we describe the changes that are needed to use AMP in the machine learning provenance and authentication scenario. 
The Content Authenticity Initiative~\cite{CAI2} is also similar to \Sys although, like AMP, CAI targets the authentication of media. Truepic is using provenance to provide a image provenance service for the insurance industry~\cite{Truepic}. Amber has also built a system for using provenance for video~\cite{amber,Newman19,AmberTalk}

The protection of the machine learning software is related to previous work in the protection of the software bill of materials (SBOM) and the software supply chain. SPDX~\cite{spdx} has been proposed as a way to protect the software bill of materials. In addition, in-toto~\cite{intotoweb,236322} provides a system for protecting the entire supply chain. 
\section{Conclusion}
\label{sec:conc}
Data poisoning attacks and mitigations are important and active areas of research, and we expect to see increasing numbers of
these types of attacks occurring in the wild in the near future. \Sys uses crytographic-based authentication and provenance to mitigate one class of data poisoning attacks where the dataset publisher can
create manifests which includes signed metadata that cryptographically binds the metadata to
the training and validation datsets used for training. 

Software poisoning attacks are another threat vector for machine learning systems, and \Sys also prevents software poisoning attacks.  Similar to the authentication of the datasets, all of the software used to train a model can also be authenticated by \Sys. 

\Sys also offers protection during inference for both the evaluation dataset, which itself may also have been signed,
as well as the authenticated machine learning model.
Once a model has been trained using signed datasets and software, it can also be signed. All of the inference software can also be verified by \Sys. 

We find that with minor modification, the AMP system, which has previously been proposed for the authentication and provenance of 
distributed media objects, can be extended to \Sys and used to provide these benefits for machine learning
datasets, software and models. 
\Sys's provenance features allow verification of all of a machine learning's subcomponents to be discovered and verified. Tracking provenance is particularly important for complex machine learning systems which require keeping track and verification of smaller machine learning subcomponents. 
As a result, a model creator can verify the training and validation sets used
during training have not been altered. Similarly, the user of a model can also authenticate both the trained model and
the evaluation set, if a manifest has also been generated for these machine learning objects during inference.


\bibliographystyle{IEEEtran}
\bibliography{vamp}

\end{document}